\documentstyle[preprint,aps]{revtex}
\draft
\preprint{SNUTP 98-137,\ KIAS-P98042}
\tightenlines
\begin{document}
\title{\Large\bf Axion and Almost Massless Quark as Ingredients
of Quintessence}
\author{Jihn E. Kim } 
\address{
Center for Theoretical Physics, Seoul National University,
Seoul 151-742, Korea, and\\
School of Physics, Korea Institute for Advanced Study, 
207-43 Cheongryangri-dong, Dongdaemun-ku, Seoul 130-012, Korea
}
\maketitle

\begin{abstract} 
It is pointed out that a natural introduction of quintessence
is an axion with an almost massless quark. With an appropriate
gauge or discrete symmetry, one can obtain the quintessence
potential height of order $(0.003\ {\rm eV})^4$. 
Even though the QCD axion with almost massless up quark is a possibility,
we stress the almost massless quark in the hidden sector
together with the QCD axion toward the quintessence.
\end{abstract}


\newpage

It has been a theoretical prejudice that the cosmological
constant is zero. But the recent Type 1a Supernova data 
strongly suggest that the cosmological constant is
nonzero at the level of $\Omega_\Lambda\simeq 0.6$ \cite{perm}.
In view of no widely accepted theory for vanishing
cosmological constant, this observation can be an experimental
hint that there cannot be a theory of  
{\it presently} vanishing cosmological constant.

Nevertheless, it is widely believed that there will be a theory
of vanishing cosmological constant where at the minimum of the potential
the vacuum energy is zero. Then the present nonvanishing cosmological
constant might be a temporary phenomenon in the sense that
the present vacuum energy happens to be close to the present 
energy density. For the temporary nonvanishing
cosmological constant, $\lq\lq$quintessence" has been proposed
\cite{cosmion}. For theoretical reasons, it
is required for this field to be a pseudo-Goldstone boson.
In this scenario, the vacuum energy density at the minimum of the
potential is zero and it 
is of order $(0.003\ {\rm eV})^4$ away from the minimum
and the mass of the pseudo-Goldstone boson is of order
$10^{-34}$ eV if the corresponding decay constant is 
the Planck scale. The most serious challenge toward this scenario
is {\it the timing problem: 
Why at this particular cosmic time is the cosmological constant
$\Lambda$ comparable to the energy density of the universe?}

In this scenario, the understanding of extremely small cosmological
constant starts from a symmetry. Of course, in the symmetric limit it
is assumed for the vacuum energy to be zero. Then, 
if this symmetry is explicitly broken, a
small cosmological constant can be generated. Since the observed
cosmological constant is extremely small compared to the other
fundamental mass scales, we must look for a mass parameter which
has a potential to be exactly zero. There are two such mass scales
among the observed particles: neutrino and up quark.
Another possible zero mass particle
is the hidden sector quarks (h-quarks) carrying hidden sector
confining color (h-color). 

In this paper, after pointing out the possibility of an extremely 
small up quark mass toward a general discussion, we present
a scenario with an almost massless h-quark(s) with axion.
It is very interesting to consider this scenario of the axion in 
the h-sector, since with the almost massless h-quark(s)
the QCD axion can become a candidate for the cold dark matter. 

A vanishing up quark mass is an attractive solution of
the strong CP problem \cite{rev}. Phenomenologically, it is known
that this possibility is not ruled out so far \cite{uquark}. For 
$m_u=0$, the chiral symmetry, 
$
u_L\rightarrow e^{i\alpha}
u_L,\ \ u_R\rightarrow e^{-i\alpha}u_R,
$
is an exact symmetry of QCD Lagrangian and the instanton potential is
not developed.  $\theta$ is not a physical parameter,
namely physical quantities should not depend on $\theta$. For 
example, for $m_u=0$ the neutron electric dipole moment
(NEDM), $d_n$, does not depend on $\theta$. In the $\theta$
vacuum, NEDM is calculated \cite{baluni,rev} to
\begin{equation}
{d_n\over e}=\xi 
{m_u \sin\theta\over f_\pi^2\sqrt{2Z\cos\theta +(1+Z^2)}}
\end{equation}
where $e$ is the electron charge, $Z=m_u/m_d$ 
and $\xi$ is a parameter of order $10^{-1}$ \cite{rev}. 
The standard limit of $|\theta|<10^{-9}$
is obtained from $m_u/m_d\sim 0.6$ and $d_n^{\rm theory}
\simeq (2-20)\times
10^{-16}\theta\ e$cm and the bound $d_n^{\rm exp}<1.2\times 10^{-25}
\ e$cm \cite{altarev}. But if $m_u=0$, $\theta$
is not bounded. It is in this sense that $\theta$ is not a
physical parameter. 

If $m_u$ is nonzero but small, then $\theta$ is physical. Nevertheless,
the NEDM bound $|d_n|<1.2\times 10^{-25}\ e $cm can be satisfied
for $\theta=$ O(1) if
\begin{equation}
m_u<2\times 10^{-13}\ {\rm GeV}.
\end{equation}  
The vacuum angle $\theta$ can be of order 1, but still the only
experimental constraint is not strong enough
to bound $\theta$. It is in this framework that vacuum energy
can be very small with very small up quark mass parameter.

Toward introducing massless up quark, 
a discrete $R_F$ parity $R_F=(-1)^{3B-L+2S+2IF}$ 
\cite{kkl} has been introduced in the supersymmetric standard model.
Here, $B, L, S, I$ are the baryon number, lepton number,
spin, and the weak isospin, respectively,
and $F$ is 1 for the first family and 0 for the others,
i.e. $F=\delta_{f1}$ where $f(=1,2,3)$ is the family number. 
The $R_F$ parity distinguishes the first family from
the second and the third families.
In this model, if the $R_F$ parity were not broken, 
the left-handed up quark
and $\nu_{eL}$ is the lightest particle with $R_F=-1$. 
Therefore, there is no particle responsible for dark matter
even in the $R_F$ conserving model, in contrast
to the theory with the conventional $R$ parity. The axion, needed
to make the height of the axion potential a tiny amount above the
true ground state, does not develop a QCD scale potential 
but a potential suppressed by $m_u/m_d\equiv Z$,
\begin{equation}
V={Z\over (1+Z)^2}f_\pi^2 m_\pi^2 \left(1-\cos{a\over F_a}\right).
\end{equation}
Therefore, the axion resulting from this theory is truly 
invisible cosmologically except for the cosmological
constant. But the astrophysical bound still applies due to
the presence of Peccei-Quinn symmetry preserving derivative
coupling $\partial^\mu a\cdot J^a_\mu$, and hence $F_a\ge
10^9$~GeV \cite{rev,turner}. Therefore, the upper bound of the axion 
mass is   
\begin{equation}
m_a\simeq \sqrt{{m_u\over m_d}}f_\pi m_\pi {10^9~{\rm GeV}\over
F_a}\le 4\times 10^{-44} \ {\rm GeV}.
\end{equation}
This extremely light axion can be called quintessence, but
cannot be the cold dark matter candidate. Furthermore, there still
exists a dispute on the viability on $m_u=0$ \cite{leut}. 

{\rm For} the axion to be the cold dark matter candidate, 
the QCD axion
with $F_a\simeq 10^{12}$~GeV which is the usual very light axion
\cite{kim}, must be saved. For this purpose, the up quark mass
is assumed to be the usual value, 5~MeV \cite{leut}.
Then the above mechanism, with an almost massless quark 
and the corresponding axion, should be placed in another sector.
The hidden sector (h-sector), 
introduced to break supersymmetry,
nicely fits to this scheme. With two nonabelian gauge groups,
we must consider two vacuum angles: $\theta$ corresponding to
QCD and $\theta_h$ corresponding to the h-color. We need two
axions: $a_1$ with the decay constant $F_1$ and $a_2$ with
the decay constant $F_2$. One combination is for the dark matter 
candidate and the other combination for the quintessence. 
Since the QCD axion is designed to be the dark matter
candidate, the quintessence potential must result from the
h-color and be very shallow. Suppose that $a_1$ and $a_2$ have
the following couplings to the QCD and the h-color field
strengths \cite{ck}
\begin{equation}
{\cal L}_{\rm eff}=\left({a_1\over F_1}+{a_2\over F_2}\right)
\{F\tilde F\}
+\left(n_1{a_1\over F_1}-n_2{a_2\over F_2}\right)\{F^\prime\tilde 
F^\prime\}
\end{equation}
where $\{F\tilde F\}\equiv (1/64\pi^2)\epsilon^{\mu\nu\rho\sigma}
F^a_{\mu\nu}F^a_{\rho\sigma}$ is the pseudoscalar density constructed
from QCD gluon field strength $F$ and similarly for the h-color
with primed field strength $F^\prime$. We have assumed the
couplings to the QCD density as $1/F_1$ and $1/F_2$
just for a simple discussion. In fact, the string compactification
example shows this behavior \cite{ck}. To have two
independent pseudo-Goldstone bosons, $n_1:n_2\ne 1:-1$
is assumed, otherwise
one Goldstone boson remains to be exactly massless and cannot
be a candidate for quintessence. The very light QCD axion is
$a=a_1\cos\gamma+a_2\sin\gamma$ where $\tan\gamma=F_1/F_2$. The QCD
axion decay constant is 
\begin{equation}
F_a={2F_1F_2\over \sqrt{F_1^2+F_2^2}}.  
\end{equation}
The other combination is the quintessence, $a_q=-a_1\sin\gamma+
a_2\cos\gamma$. Its decay constant is
\begin{equation}
F_q={\sqrt{F_1^2+F_2^2}\over n_1+n_2}.
\end{equation}

Suppose two scales $F_1$ and $F_2$ have a hierarchy, e.g. 
$F_1\ll F_2$, such as $F_2=M_P$ and $F_1=10^{12}$ GeV.
Then, $F_a\simeq 2F_1$ and $F_q\simeq F_2/(n_1+n_2)$, fulfilling
our motivation to introduce the light axion as the dark
matter particle (QCD axion with $F_a\sim 10^{12}$ GeV) and 
the remaining quintessence with the decay constant $F_q\simeq M_P$.
This becomes possible because the QCD potential is much larger
than the h-color potential. Without the almost
massless h-quark, the h-color
potential is much larger than the QCD potential and the QCD
axion decay constant is the larger one, which is the so-called
$\lq\lq$decay constant problem" of the model independent
axion \cite{pasco98}. 
 
To make the h-color potential extremely shallow, we introduce
an almost massless h-color quark. For definiteness, let us
introduce the h-color gauge group $SU(N)_h$ and six chiral h-quarks:
$\tilde U, \tilde M, \tilde D$ which are weak $SU(2)$ singlets
and $\tilde U^c, \tilde M^c, \tilde D^c$ which form a lefthanded weak 
$SU(2)$ triplet with $Y=0 $ denoted as $\tilde Q_L$,
\begin{equation}
\tilde Q_L=\left(\matrix{\tilde D^c\cr
\tilde M^c\cr
\tilde U^c}\right)_L,\ \ \tilde U_L,\ \ \tilde M_L,\ \ \tilde D_L.
\end{equation}
They transform as $(1,3,0,\bar N)$ and $(1,1,(Y=+1,0,-1),N)$ under
$SU(3)_c\times SU(2)_w\times U(1)_Y\times SU(N)_h$.
The scalar superpartners will be denoted without tilde
and subscript. Since we do
not introduce a triplet Higgs field, these h-quarks are massless
at the level of the electroweak scale.

However, these h-quarks can obtain mass suppressed by powers of $M_P$. 
Three combinations for isospin triplet operators are
the symmetric combinations of the doublet fields,
$ (H_2H_2)_{\rm sym}$, $(H_1H_2)_{\rm sym},\ (H_1H_1)_{\rm sym}
$
which carry $Y=+1,0,-1$, respectively. Without any further symmetry
imposed, therefore the h-quark masses are suppressed only by one power of
the Planck mass, which is not enough for the quintessence.
Therefore, we introduce the $Z_4$ symmetry.
The $Z_4$ charges of $H_1$ and $H_2$ are 1.
If the $Z_4$ charge of h-quarks is zero, then $d=4$ term
in the superpotential suppressed by one power of $M_P$ cannot arise. 
However, the following
nonrenormalizable terms can be present in the superpotential,
\begin{equation}
{1\over M_P^3}QM^{ij}(H_iH_j)_{\rm sym}(H_1H_2)_{\rm anti}
\end{equation}
where the symmetric combinations of $i,j$ are for 
$M^{11}=U,\ M^{10}=M,\ M^{22}=D$. 
Since this operator is suppressed by $M_P^3$ and mass 
parameters in the numerator are at the electroweak scale, we
obtain sufficiently small h-quark masses of order $10^{-42}$ GeV. 
Since the h-quark mass is extremely small, the vacuum energy
contributed by this chiral symmetry violating interaction is also
extremely small. Its magnitude is a function of
the h-quark condensate, $<\tilde U^c\tilde U>, <\tilde M^c\tilde M>$, 
and $<\tilde D^c\tilde D>$. Since the fermion 
condensate breaks supersymmetry (being
the coefficient of $\theta\theta$ term), the supersymmetric vacuum
does not allow the condensate. Thus 
we can allow its value only below the SUSY breaking scale $M_{SUSY}$.
The condensate breaks $SU(2)\times U(1)$ also. From 
the experimental bound on the $\rho$ parameter
\cite{part}, we require the
condensation scale of these h-quarks to be less than 5\% of the
electroweak scale, $<\tilde U^c\tilde U>\le 7$ GeV, etc. For 
$<\tilde U^c\tilde U>=<\tilde M^c\tilde M>=<\tilde D^c\tilde D>
\simeq 5$~GeV, we can estimate the vacuum
energy contributed by the explicit chiral symmetry breaking
interaction (9) as
\begin{equation}
V \simeq f\times 10^{-43}\tan^2\beta\cos^4\beta\ {\rm GeV}^4
\end{equation} 
where $f$ is a coupling constant potentially arising 
in this estimation. For $\tan\beta=2$ and 30, $f$ of order $10^{-3}$ 
and 1, respectively, give the vacuum energy of order $10^{-47}$~GeV.

The other chiral symmetry breaking h-color instanton potential is
\begin{equation}
V\sim {1\over \Lambda_h^{N-1}}
{m_{\tilde U}m_{\tilde M}m_{\tilde D}M_{\rm h-gluino}^N}\left(
1-\cos{a_q\over F_q}\right)
\end{equation}
where $\Lambda_h\sim 10^{13}$ GeV is the confining scale of $SU(N)_h$.
Since Eq. (11) gives much smaller contribution to the vacuum energy
for any $N$,
Eq. (10) is the dominant vacuum energy. Because Eq. (11) is not
the dominant potential, $\theta_h\ne 0$ at the minimum of
the potential. Nevertheless, the QCD vacuum angle $\theta$ is very
close to zero since the QCD instanton interaction of order $F_\pi^2
m_\pi^2$ is the dominant one for $\theta$.  

{\rm For} $F_q$ of order $10^{16}$ GeV \cite{mia},
the quintessence mass is estimated as 
$10^{-40}$ GeV. On the other hand, if we use $F_q=M_P$,
the mass will be about $10^{-43}$ GeV. 

In passing, we point out that in models without h-quarks \cite{gkn}
we consider only the chiral symmetry for the h-gluino
of $SU(N)_h$.  The instanton contribution to the potential 
is the form given by Eq.~(11) without
h-quark mass terms and the corresponding suppression factor with
$\Lambda_h$. Since the h-gluino mass scale is of order $M_{SUSY}\sim
\Lambda_h^3/M_P^2$, the height of the potential is of order
$\Lambda_h^{2N+4}/M_P^{2N}$. For a $Z_D$ discrete symmetry subgroup
of $U(1)_R$ \cite{gkn}
where $D$ is a positive even integer, the
explicit chiral symmetry breaking term is $(1/M_P^{3D/2-3})
\int d^2\theta (W^\alpha W_\alpha)^{D/2}$. 
This explicit chiral symmetry breaking term is more
important than Eq.~(11) for $N>(1/4)(3D-2)$. 
It is known that $D\ge 6$ gives
the height less than $10^{-13}\ {\rm GeV}^4$ \cite{gkn}. 
To realize the quintessence from
this h-gluino chiral symmetry, we need $D\sim 10$ and
$N>(1/4)(3D-2)$. For example, it is satisfied for $Z_5$
\cite{Z5} and $N\ge 8$.
Then the quintessence is composite and arising from the h-gluino 
condensation. The other possibility that Eq.~(11) is
more important than the explicit h-gluino chiral 
symmetry breaking term is not elaborated here.

If the h-quarks are responsible for the shallow potential
as discussed above, the lightest supersymmetric 
particle (LSP) can be the cold dark matter candidate also. 
But to solve the strong CP problem, we need a QCD axion.

In conclusion, we have explored the possibility of the quintessence
being an axion in the hidden sector. However, this h-sector
axion has an extremely shallow potential due to almost massless
h-quarks. Even being an axion, the hidden-sector vacuum angle $\theta_h$
at the minimum of the potential is nonzero due to the explicit
breaking of the Peccei-Quinn symmetry of the h-sector.
The timing problem is solved by the discrete $Z_4$ symmetry
to produce a dominant effective superpotential suppressed by $M_P^3$. 
The cold dark matter needed for large scale structure formation 
is the conventional very light axion \cite{kim} or the LSP. 
The model presented in this paper is just one example among
multitude of models fulfilling almost vanishing h-sector potentials.

\acknowledgments
This work is supported in part by KOSEF,  MOE through
BSRI 98-2468, and Korea Research Foundation.


\begin{references}

\bibitem{perm}S. Perlmutter {\it et al.}, Astrophys. J.
{\bf 483}, 565 (1997); S. Perlmutter {\it et al.}, B.A.A.S. {\bf 5},
1351 (1998); A. Riess {\it et al.}, Astrophys. J., in press (1998); 
S. Perlmutter, talk presented at COSMO-98,
Asilomar, Monterey, CA, Nov. 16--20, 1998.

\bibitem{cosmion} M. Bronstein, Physikalische Zeitschrift Sowjet Union
{\bf 3}, 73 (1933); M. ${\ddot {\rm O}}$zer 
and M. O. Taha, Nucl. Phys. {\bf B287},
797(1987); B. Ratra and P. J. E. Peebles, Phys. Rev. {\bf D37},
3406 (1988); J. A. Frieman, C. T. Hill and R. Watkins,
Phys. Rev. {\bf D46}, 1226 (1992); R. Caldwell, R. Dave and P. J.
Steinhardt, Phys. Rev. Lett. {\bf 80}, 1582 (1998).

\bibitem{rev} J. E. Kim, Phys. Rep. {\bf 150}, 1 (1987);
H. Y. Cheng, Phys. Rep. {\bf 158}, 1 (1988); R. D. Peccei,
in {\it CP Violation}, ed. C. Jarlskog (World Scientific,
1989) p.503.

\bibitem{uquark} H. Georgi and I. N. McArthur, Harvard Univ. 
Report. HUTP-81/A011 (1981);
D. B. Kaplan and A. Manohar, 
Phys. Rev. Lett. {\bf 56}, 2004 (1986); K. Choi,
C. W. Kim and W. K. Sze, Phys. Rev. Lett. {\bf 61}, 794 (1988).
But this issue is not settled yet, as discussed in
H. Leutwyler, Phys. Lett. {\bf B378}, 313 (1996).

\bibitem{leut} H. Leutwyler, Ref.~\cite{uquark}.

\bibitem{baluni} V. Baluni, Phys. Rev. {\bf D19}, 2227 (1979).

\bibitem{altarev} I. S. Altarev {\it et al.}, 
Phys. Lett. {\bf B276}, 242 (1992).

\bibitem{kkl} J. E. Kim, B. Kyae and J. S. Lee, KIAS preprint
KIAS-P98041 (1998) [hep-ph/9811510].

\bibitem{mu} J. E. Kim and H. P. Nilles, Phys. Lett. {\bf B138},
150 (1984).

\bibitem{kn} J. E. Kim and H. P. Nilles, Mod. Phys. Lett.
{\bf A9}, 3575 (1994).

\bibitem{turner} For the supernova emmision of very light axions,
see, A. Pantziris and K. Kang, Phys. Rev. {\bf D33},
3509 (1986), and {\bf D43}, 241 (1991); 
G. Raffelt and D. Seckel, Phys. Rev. Lett. {\bf 60}, 1793 (1988); 
M. S. Turner, {\it ibid} {\bf 60}, 1797 (1988); R. Mayle {\it
et al.} Phys. Lett. {\bf B203}, 188 (1988).

\bibitem{kim} J. E. Kim, Phys. Rev. Lett. {\bf 43}, 103 (1979);
M. A. Shifman, V. I. Vainstein and V. I. Zakharov,
Nucl. Phys. {\bf B 166}, 4933 (1980);
M. Dine, W. Fischler and M. Srednicki, Phys. Lett. {\bf B104},
199 (1981); A. Zhitniskii, Sov. J. Nucl. Phys. {\bf 31},
260 (1980).

\bibitem{ck} For an example, see, K. Choi and J. E. Kim, 
Phys. Lett. {\bf B165}, 71 (1985).

\bibitem{pasco98} J. E. Kim, talk presented at PASCO-98, 
hep-ph/9807322.

\bibitem{part} Particle data group, European Phys. J. {\bf 3}, 97 (1998).

\bibitem{mia} K. Choi and J. E. Kim, Phys. Lett. {\bf B154},
393 (1985).

\bibitem{gkn} H. Georgi, J. E. Kim and H. P. Nilles,
Phys. Lett. {\bf B437}, 325 (1998) [hep-ph/9805503].

\bibitem{Z5} Since we must consider h-gluino bilinears, $(W^\alpha
W_\alpha)^{5}$ is the lowest dimension operator
breaking $U(1)_R$.

\end{references}
\end{document}